\def\year{2023}\relax

\documentclass[letterpaper]{article} 
\usepackage{aaai23}  
\usepackage{times}  
\usepackage{helvet}  
\usepackage{courier}  
\usepackage[hyphens]{url}  
\usepackage{graphicx} 
\usepackage{natbib}  
\usepackage{caption} 
\usepackage{algorithm}
\usepackage{algorithmic}

\usepackage{newfloat}
\usepackage{listings}
\DeclareCaptionStyle{ruled}{labelfont=normalfont,labelsep=colon,strut=off} 
\floatstyle{ruled}
\newfloat{listing}{tb}{lst}{}
\floatname{listing}{Listing}
\usepackage{subcaption}
\usepackage{tabularx}
\usepackage{booktabs}
\usepackage{subcaption}
\usepackage{tabularx}
\usepackage{booktabs}
\usepackage{comment}
\usepackage{multirow}
\usepackage{amsmath}
\usepackage{CJKutf8}
\usepackage{subcaption}
\usepackage{tabularx}
\usepackage{booktabs}
\usepackage{comment}
\usepackage{multirow}
\usepackage{amsmath}
\usepackage{soul}
\usepackage{xspace}
\usepackage{xcolor,colortbl}

\usepackage{xcolor}
\usepackage{xspace}
\usepackage{soul}

\makeatletter
\g@addto@macro{\endtabular}{\gdef\rowfonttype{}}
\makeatother
\newcommand{\rowfonttype}{}
\newcolumntype{L}{>{\rowfonttype\strut}l}
\title{``A Special Operation'': A Quantitative Approach to Dissecting and Comparing Different Media Ecosystems' Coverage of the Russo-Ukrainian War}
\author {
    Hans W.A. Hanley, 
    Deepak Kumar, 
    Zakir Durumeric\\
}
\affiliations {
    Stanford University \\
   
    hhanley@stanford.edu, kumarde@stanford.edu, zakird@stanford.edu
}

\begin{document}

\maketitle

\begin{abstract}
The coverage of the Russian invasion of Ukraine has varied widely between Western, Russian, and Chinese media ecosystems with propaganda, disinformation, and narrative spins present in all three. By utilizing the normalized pointwise mutual information metric, differential sentiment analysis, word2vec models, and partially labeled Dirichlet allocation, we present a quantitative analysis of the differences in coverage amongst these three news ecosystems. We find that while the Western press outlets have focused on the military and humanitarian aspects of the war, Russian media have focused on the purported justifications for the ``special military operation'' such as the presence in Ukraine of ``bio-weapons'' and ``neo-nazis'', and Chinese news media have concentrated on the conflict's diplomatic and economic consequences. Detecting the presence of several Russian disinformation narratives in the articles of several Chinese media outlets, we finally measure the degree to which Russian media has influenced Chinese coverage across Chinese outlets' news articles, Weibo accounts, and Twitter accounts. Our analysis indicates that since the Russian invasion of Ukraine, Chinese state media outlets have increasingly cited Russian outlets as news sources and spread Russian disinformation narratives.
\end{abstract}

\section{Introduction}
On February 24, 2022, Russian Federation President Vladimir Putin announced a ``special military operation to demilitarize and denazify'' Ukraine~\cite{Thompson2022}. Following the initial invasion, media outlets in various parts of the world covered the war in drastically different lights. For instance, the Western press (\textit{e.g.}, CNN, Fox News, New York Times) labeled the ``special operation'' a ``war crime'' laden ``unprovoked invasion'' perpetrated by the Russian government~\cite{Phillip2022}. Russian outlets (\textit{e.g.}, Sputnik News, Russia Today), in turn, have largely denied any war crimes, placing fault for the necessity of the ``special operation'' on Western countries~\cite{Thompson2022}. Chinese state media outlets (\textit{e.g.}, China Today, People's Daily) meanwhile have advocated for diplomacy while simultaneously blaming Western powers for sparking the conflict. However, despite the evident differences and the large impact these differences have had on different populations' perceptions of the war, there has not been a systematic analysis of the narratives present in these three media ecosystems\cite{McCarthy2022}. In this work, we present one of the first quantitative analyses of the key differences and similarities between the narratives touted by Chinese, Russian, and Western outlets about the Russo-Ukrainian War. Specifically, we propose an approach to determine and compare \textit{which} topics media ecosystems report on, \textit{how} each ecosystem reported on those topics, and finally \textit{whether} there was influence between different media ecosystems (\textit{e.g.}, Russian to Chinese). 


To perform our analysis, we first curate a dataset of 11,359 different articles about Ukraine from Western (4,536~articles from eight outlets), Russian (3,572~articles from ten outlets), and Chinese (3,251~articles from seven outlets) news ecosystems. Utilizing differential sentiment analysis and word2vec models, we then detail how each ecosystem has largely covered aspects of the Russo-Ukrainian War. Using a normalized scaled pointwise mutual information metric (NPMI) and partially labeled Dirichlet allocation (PLDA), we then extract the most characteristic words and topics from each ecosystem's articles. We quantitatively show that while Western outlets have persistently labeled the ``invasion'' as a ``war'' and described the ``crimes'' committed throughout Ukraine, \textit{both} Russian and Chinese news outlets have largely characterized the invasion as a ``crisis'' or a ``conflict.'' Similarly, while the Western press has focused on the humanitarian and day-to-day military aspects of the war, Russian outlets have focused on justifications for the ``special military operations'' like the presence of ``bio-weapons'' and ``neo-nazis'' in Ukraine and Chinese news outlets have concentrated on the diplomatic and economic fallout of the invasion.

\begin{table*}
\centering
\fontsize{9pt}{8.5pt}
\selectfont
\setlength{\tabcolsep}{4pt}
\begin{tabular}{lrlrlr}
\multicolumn{2}{c}{{\textbf{Western} }}   & \multicolumn{2}{c}{{\textbf{Russian}}}  & \multicolumn{2}{c}{{\textbf{Chinese} } }
\\ \cmidrule(lr){1-2} \cmidrule(lr){3-4} \cmidrule(lr){5-6}
  {Domain} & {Articles} &{Domain} & {Articles} & {Domain} & {Articles}           \\   \midrule
  cnn.com  & 477 & tass.com  & 861   &  chinadaily.com.cn   &   1044            \\ 
 nytimes.com& 571 & sputniknews.com & 743   &   cgtn.com     &          966     \\ 
  washingtonpost.com  & 414  &news-front.info   & 508   &  globaltimes.cn  &     367            \\ 
 theguardian.com& 636 &geopolitica.ru & 80   & ecns.cn  & 702 \\   
  yahoo.com & 774 & southfront.org  & 241  &  xinhuanet.com   &      430          \\ 
 reuters.com & 716 & katehon.com & 65   & pdnews.cn       &  286     \\  
 foxnews.com & 566 & journal-neo.org & 90   & english.cctv.com     &   107           \\  
  nbcnews.com  & 382 & rt.com  & 689   &  --      & --            \\ 
  --  & -- & strategic-culture.org  & 102   &  --         &--         \\ 
  --   & -- & waronfakes.com  & 193   &  --          & --         \\ \midrule
 \multicolumn{1}{l}{\textbf{Total} }  & 4536&  &3572 & & 3251  \\ 
\end{tabular}
\caption{\label{tab:news-websites} We gather a set of English-language articles about Ukraine from Western, Russian, and Chinese media ecosystems. }  
\vspace{-10pt}
\end{table*}
After performing our comparative topic analysis, we identified the repeated presence of several Russian disinformation narratives~\cite{Price2022}\, within the Chinese news ecosystem, particularly about US-funded Ukrainian biological weapons facilities. We thus measure the degree to which Russian news outlets have influenced Chinese news outlets' coverage. Specifically, we document the frequency that seven different Chinese outlets use Russian news outlets as sources and their reuse of Russian-sourced images within their coverage of the war on their websites, Twitter accounts, and Weibo accounts (a Chinese version of Twitter). Observing a marked increase in Chinese state media citations of Russian sources beginning in early February 2022, we finally measure how an extended group of 39~Chinese media outlets interacted with and promoted Russian disinformation narratives on Weibo. Looking at the popularity of Chinese news outlets' posts about Russian disinformation topics on Weibo, we find that these posts enjoyed higher levels of popularity compared to posts that do not reference these disinformation stories. Finally, fitting a negative binomial regression to model the number of Weibo posts from Chinese news outlets about different Russian disinformation campaigns, we find that as Chinese news outlets cite more Russian outlets as news sources, they are more likely to post disinformation. 

Our work underscores the importance of performing analyses across multiple platforms and media ecosystems in understanding the nuances of how global events are framed, how different populations interpret and digest world events, and how disinformation originates and spreads.
The Russo-Ukrainian War is a \textit{global} event with \textit{global} implications every country must consider ranging from skyrocketing international food prices, the resettlement of refugees, and threats of nuclear fallout~\cite{Treisman2022,Thompson2022}; focusing only on news and campaigns targeted at Western to understand how populations are processing these implications can only go so far. We hope our quantitative approach can serve as the basis for future studies. 



\section{Methodology}\label{sec:methodology}

To perform our comparative analysis of the attitudes, narratives, and topics discussed by the Western press, Chinese state media, and Russian propaganda websites, we collect a total of 11,359  unique news articles published between January 1, 2022, and April 15, 2022 (Table~\ref{tab:news-websites}). To later understand the degree of Russian influence on Chinese media, we collect the social media feeds of major Chinese state media outlets and Russian state actors on both Weibo and Twitter.

\vspace{2pt} \noindent
\textbf{News Articles.} Our news article dataset consists of published pieces from Western news websites, English-language Russian websites, and English-language Chinese websites (Table~\ref{tab:news-websites}). For lack of a better term, we use the term ``Western'' to describe press widely circulated in the global ``West'' (\textit{e.g.}, US, UK)~\cite{WesternPopulation}. We refer to websites as ``Russian'' if they are Russian state media, were identified as ``proxies'' for the Russian government, or are Russian propaganda~\cite{RussiaPillar2020}. Lastly, we refer to websites as ``Chinese'' if they are Chinese state media outlets. 

For our list of Western outlets, we manually selected eight highly popular mainstream news websites from across the political spectrum~\cite{zannettou2017web}. In addition to a set of nine Russian websites identified by the US State Department~\cite{RussiaPillar2020}, for our Russian dataset, we include the recently launched waronfakes.com. Since its initial appearance in March 2022, the New York Times and others have investigated the site as a hub of Russian disinformation~\cite{Thompson2022,hanley2022happenstance}. For our list of Chinese media news websites, we utilize seven English-language news websites identified by the US State Department as Chinese ``foreign missions''~\cite{ortagus2020designation}.
We recognize that these lists do not incorporate \textit{all} articles circulated in each media ecosystem and thus are naturally biased. However, our selection of websites \textit{do} represent a cross-section of some of the most widely circulated news sources in each ecosystem and thus provide indications of reporting for Western~\cite{YouGov}, Russian~\cite{RussiaPillar2020}, and Chinese~\cite{ortagus2020designation} news media.

We utilize a breadth-first scraping algorithm and the Python \texttt{Selenium} package to collect the set of English-language articles that each of our websites published about Ukraine. Specifically, for each website, we scrape 5~hops from the root page (\textit{i.e.}, we collect all URLs linked from the homepage [1st hop], then all URLs linked from those pages [2nd hop], and so forth). To get Ukraine-related articles, for each website page, we use the Python \texttt{newspaker3k} library to collect article contents and to determine if the article mentions ``Ukraine''. We further supplement this corpus by using Google's API to find and add articles indexed in 2022 that mention Ukraine. We note that due to the lack of precision in acquiring the publication date of each article with \texttt{newspaker3k}, we utilize the Python library \texttt{htmldate} to extract each article's publish date. Altogether, between January 1, 2022, and April 15, 2022, we collect 11,359~articles about Ukraine; 4,536~from Western Press outlets, 3,572~from Russian propaganda websites, and 3,251~from Chinese state media (Table~\ref{tab:news-websites}).

\vspace{2pt} \noindent
\textbf{Weibo Dataset.}
To understand the degree of Russian influence on Chinese media reports and discussions surrounding the Russo-Ukrainian War, we also collect posts from Weibo, a Chinese Mandarin-language version of Twitter~\cite{McCarthy2022}.\begin{CJK*}{UTF8}{gkai}  
We collect the posts of the accounts of the seven different Chinese state media organizations from our news article dataset (for the CGTN news organization, we collect the Weibo posts of its CGTN and CGTN journalist group/CGTN记者团 accounts). To help quantify the connection of each of these media organizations to the Russian government and Russian state media, we further scrape the accounts of the Russian Embassy/@俄罗斯驻华大使馆, Russia Today/@今日俄罗斯RT, and Sputnik News/@俄罗斯卫星通讯社. Lastly, in addition to our Chinese news organizations' Weibo accounts and Russian state-sponsored Weibo accounts, we collect the Weibo posts of the 200 users who most prominently discussed the Russo-Ukrainian conflict at the end of February as labeled by Fung et al.~\cite{fung2022weibo}. This list was manually created from users who ``actively posted about and ranked among the top posts of trending hashtags related to the Russo-Ukrainian war.'' 
After combining our lists of Weibo users, and removing inactive and duplicate accounts, we had a total of 191 distinct accounts. For each account in our dataset, we scraped the account on four occasions (March 14, March 28, April 06, and April 16) to ensure our dataset was comprehensive. To scrape each Weibo account, we utilize the Python \texttt{weibo-scraper} tool.\footnote{\url{https://github.com/Xarrow/weibo-scraper}} Ultimately, our dataset consists of 191~different accounts and 343,435~distinct Weibo posts from between January 1 and April 15, 2022.
\end{CJK*}
\begin{table*}
\centering
\centering
\fontsize{9pt}{8.5pt}
\selectfont
\setlength{\tabcolsep}{4pt}
\begin{tabular}{c|lrrr}

\toprule

  &   & Western & Russian & Chinese  \\
Topic& Keywords& {Articles} & {Articles} & { Articles}
\\ \midrule
   1  &  border,troops,exercises,belarus,eastern,invade,tensions,imminent,attack,borders

     &\textbf{137} (3.0\%)   &\textbf{90} (2.5\%)&70 (2.2\%) \\ 
  2  & sanctions,biden,tuesday,thursday,statement,address,announced,meeting,impose,united

 & 81 (1.7\%) &43 (1.2\%) & \textbf{91} (2.8\%)\\ 
   3  & city,shelling,killed,forces,destroyed,reported,residents,building,southern,near

 &74 (1.6\%)    &  63 (1.8\%) &60 (1.8\%)\\ 
  4&  refugees,refugee,poland,million,children,fleeing,ukrainians,fled,border,polish

& 98 (2.1\%) & 22 (0.6\%) &69 (2.1\%) \\   
   5  & nato,alliance,europe,expansion,security,soviet,west,treaty,european,eastward
 &47 (1.0\%) &  81 (2.3\%) &58 (1.8\%)  \\ 
\end{tabular}
\caption{\label{tab:shared-topics} The top five shared latent topics (in terms of the number of articles) in the Western, Russian, and Chinese ecosystems. }  
\vspace{-10pt}
\end{table*}

\vspace{2pt} \noindent
\textbf{Twitter Dataset.}
In addition to our Weibo dataset, we further collect the tweets of the seven different news Chinese news outlets within our news article dataset (China Daily, CGTN, Global Times, Chinese News Service, Xin Hua, People's Daily, and CCTV). Unlike for our Weibo dataset, we do not collect the set of Chinese users who most prominently discussed the Russo-Ukrainian conflict on Twitter (Twitter has been banned in China since 2009~\cite{Barry2022}), limiting our Twitter analysis to these seven major state-sponsored Chinese outlets who also regularly tweet. To investigate these accounts' connection to the Russian government and Russian news media, we again collect the tweets of the RussianEmbassy/@RussianEmbassy, Russia Today/@RT\_com, and Sputnik News/@SputnikInt. We collect the tweets of each account using the Tweepy API~\cite{roesslein2009tweepy} on four different instances (March 06, March 13, April 02, and April 16). Ultimately, our Twitter dataset consists of 62,717 unique tweets from 10 different accounts from January 1 and April 15, 2022.

\vspace{2pt} \noindent
\textbf{Pointwise Mutual Information.}
To determine different news ecosystems' associations with distinct words, we utilize the normalized pointwise mutual information metric. Pointwise mutual information (PMI) is an information-theoretic measure for discovering associations amongst words~\cite{bouma2009normalized}. However, as in Kessler et~al., rather than finding the pointwise mutual information between different words, we utilize this measure to understand words' association with different categories~\cite{kessler2017scattertext}. In this way, we seek to identify the characteristic words of each ecosystem's coverage of the Russo-Ukrainian War (\textit{i.e.}, Western, Chinese, and Russian media). We utilize the normalized and scaled version of PMI to prevent our metric from being biased towards rarely occurring words and to increase interpretability. Scaled normalized PMI (NMPI) for a $word_i$ and each category $C_j$ is calculated as follows:
\begin{align*}
\scriptsize
PMI(word_i, C_j) = log_2\frac{P(word_i,C_j)}{P(word_i) P(c_i)} \\
NPMI(word_i, C_j) =\frac{ PMI(word_i, C_j)}{-log_2(P(word_i, C_j))}
\end{align*}
where $P$ is the probability of occurrence and a scaling parameter $\alpha$ is added to the counts of each word. NPMI ranges between (-1,1). We choose $\alpha = 50$ given the size of our dataset~\cite{turney2001mining}. An NPMI value of \,$-1$ represents that the word and the category never occur together (given that we utilize the scaled version this never occurs), 0 represents independence, and +1 represents perfect co-occurrence~\cite{bouma2009normalized}. Finally, before computing NMPI on our dataset, we first lemmatize and remove stop words as in prior work~\cite{zannettou2020quantitative}.

\vspace{2pt} \noindent
\textbf{Partially Labelled Latent Dirichlet Allocation.} In addition to identifying words characteristic of each news ecosystem, we also extract the set of topics that are distinctive to each ecosystem. To do this, we utilize Partially Labelled Dirichlet Allocation (PLDA). PLDA is an extension of the widely-used topic analysis algorithm Latent Dirichlet Allocation (LDA)~\cite{ramage2011partially}. PLDA, like LDA, assumes that each document is composed of a distribution of different topics (which themselves are composed as a distribution of different words). However, unlike LDA, each document can form topics from a pool associated with one or more of its specific labels. For example, a newspaper article from nytimes.com, which is labeled as ``Western'', can draw from a set of labeled topics associated with ``Western'' (as opposed to an article from chinadaily.com.cn which can draw from a set of labeled topics associated with ``Chinese''). In addition to drawing from the distribution of topics associated with its labels, documents also further draw from a pool of \textit{latent} topics that are associated with every document in the dataset. PLDA can thus model the topics that are common to every document while also identifying discriminating topics for each label (\textit{i.e.}, topics specific to ``Western'', ``Russian'', ``Chinese'').

Again when fitting our PLDA model, we first lemmatize and remove stop words. When computing topics, we further weight words using term-frequency inverse document frequency (TF-IDF). Previous work has shown that this weighting leads to more accurate topics~\cite{zannettou2020quantitative}. To find the appropriate amount of topics, we optimize the word2vec topic coherence score $c_v$ that measures the semantic similarity among extracted topic words~\cite{zannettou2020quantitative}. We utilize a baseline number of 300 latent topics, varying the number of topics per label from 1 to 20. We achieve the best coherence score of 0.46 with 15 topics associated with each label (345 total topics).

\section{Comparing  The Coverage of the War}
\label{sec:compare_text}

In this section, we perform a quantitative study of the shared and distinctive narratives and topics discussed within the Russian, Chinese, and Western ecosystems.

\vspace{-5pt}
\subsection{Media Ecosystems' Shared Topics}
We begin our analysis by highlighting the topics (Table~\ref{tab:shared-topics}), as elicited by our PLDA model, that are shared between our three distinct media ecosystems from articles published between January 1 and April 15, 2022.
\begin{figure}
\centering
\begin{subfigure}{.46\textwidth}
\centering
  \includegraphics[width=0.76\linewidth]{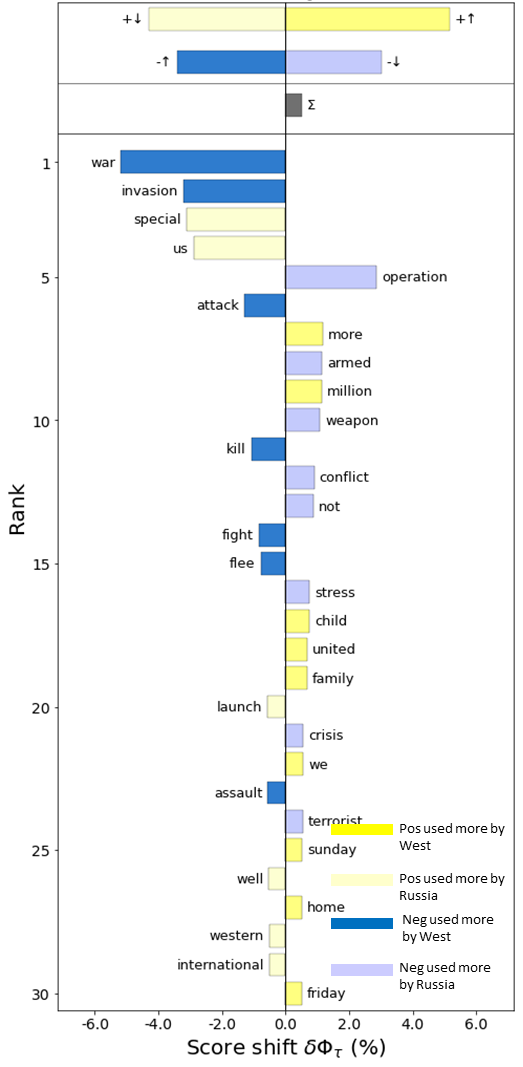} 

\end{subfigure}%
 \caption{Western vs Russian Sentiment.
  }
\label{fig:ukraine_western_russian}
\vspace{-15pt}
\end{figure}

The most common shared topic---by article count---refers to military activities in the build-up to the war (137~western articles, 90~Russian articles, 70~Chinese articles). Starting in the Spring of 2021 and continuing into February of 2022, Russia began to amass troops on their shared border with Ukraine~\cite{Schmollinger2022}. Each ecosystem wrote a correspondingly large amount about the growth of the Russian military presence. Each ecosystem further discusses the fallout following the invasion. 214 different articles (Topic 2) focused on US and EU sanctions placed on Russia (81~western articles, 43~Russian articles, 91~Chinese articles).
\begin{figure}
 \centering
\begin{subfigure}{.46\textwidth}
  \centering
  \includegraphics[width=.76\linewidth]{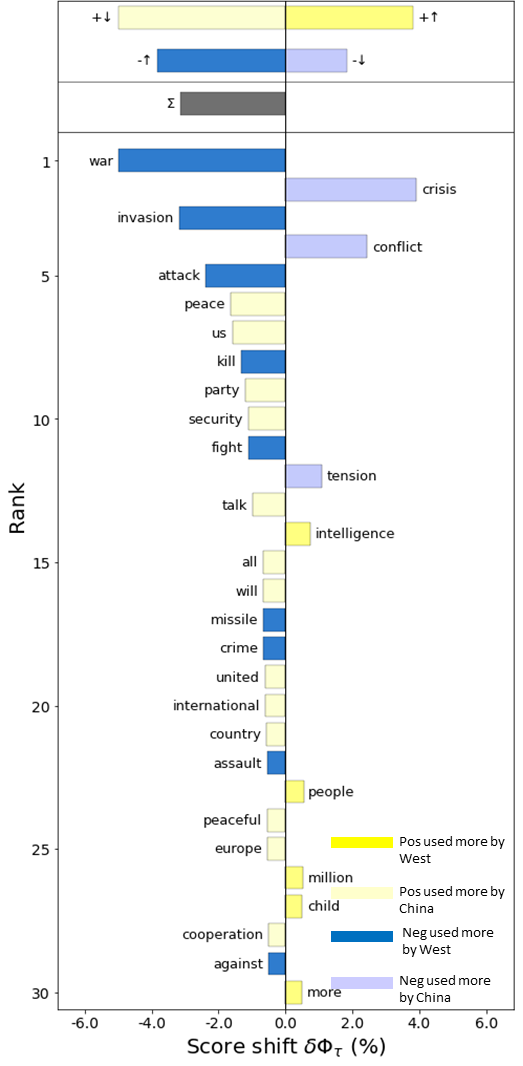}

\end{subfigure}
  \caption{Western vs Chinese Sentiment.}
  \label{fig:ukraine_western_chinese}
  \vspace{-15pt}
\end{figure}
Each ecosystem also discusses the North Atlantic Treaty Organization's (NATO's) role in the conflict (47~western articles, 81~Russian articles, 58~Chinese articles), with Russian articles, in particular, emphasizing its role. The North Atlantic Treaty Organization/NATO was founded in 1949 as a defensive military alliance between the United States, Canada, and several Western European countries against the threat of the Soviet Union. The organization adopted the agreement, called Article 5, that an attack on one member of the alliance would be treated as an attack on all members~\cite{NATO}. Since the fall of the Soviet Union, several new countries have joined the defensive pact including Hungary, Poland, and the Czech Republic, with North Macedonia being the newest member (2020), leading to increased tensions with Russia~\cite{NATO}. 

Finally, each ecosystem has written extensively about the Ukrainian refugee crisis, with 98~western articles, 22~Russian articles, and 58~Chinese articles documenting the displacement of more than 6 million Ukrainians~\cite{Semczuk2022}.

 \begin{figure}
 \centering
\begin{subfigure}{.46\textwidth}
  \centering
  \includegraphics[width=.76\linewidth]{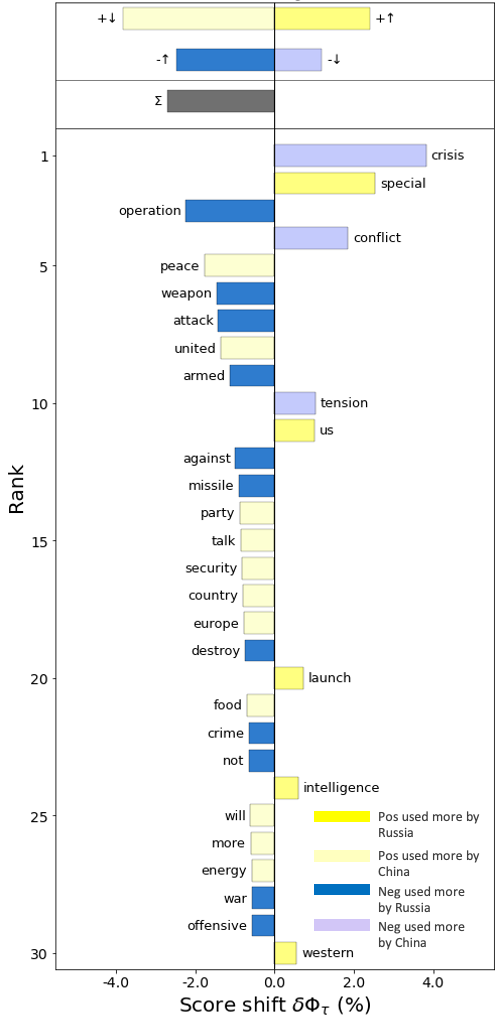}
\end{subfigure}
 \caption{Russian vs Chinese Sentiment.}
  \label{fig:ukraine_chinese_russian}
  \vspace{-15pt}
\end{figure}

\noindent \textbf{Comparing Shared Coverage with Differential Sentiment Analysis.}
We note that despite having many articles that share topics, each ecosystem often described these topics differently. To capture how each ecosystem generally wrote about the war, we thus run sentiment analysis using the labMT sentiment analysis toolkit~\cite{dodds2011temporal}. labMT grades individual words on a scale from 1 (most negative) to 9 (most positive).

We find that each ecosystem wrote about the Russo-Ukrainian War somewhat neutrally. Western coverage was the most negative with an average article sentiment of 5.42; Russian coverage came next at 5.43 and Chinese coverage was the most positive at 5.62. We confirm this ranking by further extracting the average article sentiment using the Vader sentiment Python library (scores on a scale of -1 [most negative] to +1 [most positive]) getting scores of -0.089, -0.072, and 0.017 for the Western, Russian, and Chinese ecosystems respectively~\cite{hutto2014vader}.

To help explain the differences in sentiment scores between individual ecosystems, we next identify which negatively and positively polarized words contributed the most to each dataset's sentiment score; we visualize this in pairwise graphs in Figures~\ref{fig:ukraine_western_russian},~\ref{fig:ukraine_western_chinese},~\ref{fig:ukraine_chinese_russian}. These graphs compare the relative shift in the labMT sentiment score caused by the frequency of negatively and positively polarized words.\footnote{Dark blue (negative words) and dark yellow (positive words) indicate words used more often by Western articles. Light blue (negative words) and light yellow (positive words) indicate words used more often by Russian articles. Made with~\cite{gallagher2021generalized}.}
Looking at the contribution of words to each ecosystem's sentiment score, we see that the  Western media's consistent reference to the Russo-Ukrainian conflict as a ``war'' and an ``invasion'' contributes the most to its relative negativity (Figures~\ref{fig:ukraine_western_russian} and~\ref{fig:ukraine_western_chinese}). In contrast, the Russian ecosystem,  largely calls the Russo-Ukrainian War a ``special operation'', a ``conflict'', or a ``crisis'' (Figure~\ref{fig:ukraine_western_russian}).

The Western usage of more violent and negative language is further mirrored in differential sentiment between the Western press and Chinese media (Figure~\ref{fig:ukraine_western_chinese}). The Western press commonly refers to ``attacks'', ``killing'', ``crimes'', ``assaults'', and ``refugees.'' The Chinese media ecosystem, in contrast, is more likely to write about prospects of ``peace'', talk about ``security'', or suggest ``cooperation'' (Figure~\ref{fig:ukraine_western_chinese}). Furthermore, the Chinese media ecosystem has largely avoided using harsh language to describe the Russo-Ukrainian War, using words like ``war'', ``attacks'',  and ``invasion'', even less frequently than in the Russian media ecosystem (Figure~\ref{fig:ukraine_chinese_russian}). This largely explains why the Chinese media ecosystem remained more positive than both the Western and Russian media, despite all outlets discussing similar war-related topics.

\begin{table*}
\centering
\centering
\fontsize{9pt}{8.5pt}
\selectfont
\setlength{\tabcolsep}{4pt}
\begin{tabular}{c|lll c| |lll c| |lll }

\toprule
\multirow{2}{*}{Rank}  & \multicolumn{3}{c}{\textbf{Bucha}} & & \multicolumn{3}{c}{\textbf{Mariupol}}&& \multicolumn{3}{c}{\textbf{Ukraine}} \\ \cmidrule{2-4} \cmidrule{6-8} \cmidrule{10-12}
   & {Western} & {Russian} & {Chinese} &&{Western} & {Russian} & {Chinese}     &&{Western} & {Russian} & {Chinese}        \\  \midrule

   1  & atrocity & massacre  & killing  & &  besiege   &  volnovakha     &   volnovakha &&  country & country  & underway \\ 
  2  & slay & killing & atrocity  & &   city   &   city    &    besiege &&    prepare & aim   & postpone    \\ 
   3  & suburb & atrocity  & grave   &&  devastated &   kharkiv  &   shelling&& russia    &  demilitarize &disarm   \\ 
  4& gruesome &mayor & exhume   && port  & sumy           & bombardment &&    continue&   creation &  russia \\   
   5  & killing & town  & disturbing   &&  besieged   &   liberate  &mykolaiv  &&  rearm & denazify  &     dangerous    \\ 
   6  & massacre & footage  & incident   &&  boychenko   & azov       &   kherson&& hasten  &   kiev  &  minimal\\ 
   7  & body & body  & dead  & &  boichenko   &  blockade   &    melitopol&& impending & wake   &   calm  \\
   8  & grisly & image  & scene   &&  siege   & afu        &  berdyansk&& monthslong &prompt  &    continue   \\ 
   9  & discover & dead  & massacre   &&  trap   &  militia      &  rubble&&  unabated& pledge  &  regard     \\ 
   10  & anatoliy& alleged  & anatoly  & &  volnovakha   &  surround &      northeastern && necessitate &russia    &   flare    \\
   11  & fedoruk & evidence  & strew   &&  berdiansk   &   corridor  &   theater  &&precursor& demilitarise &   prompt \\ 
   12  & corpse & murder  & mayor   &&  kherson   &   retreat    &  boychenko &&ready & military&  swift\\
   13  & sickening & torture  & torture   &&  mangush   &  hospital   &     sumy &&    enduring & ongoing &  mount \\ 
   14  & unspeakable & accusation  & alleged  & &  vershynin   & militant    &     odesa &&  bolstered & seek & blitz      \\ 
   15  & unearth & staged  & image   &&  berdyansk   &  kharkov     &    energodar&&   besides& february    &  formal\\
   16  & horrifying & picture  & theater   &&  encircle   &  regiment  &      tokmak && abet  & moscow  & official\\
   17  & hostomel & fake  & staged   &&  remained   &  village      &  southeastern && estimation&argue &   assure \\ 
   18  & borodyanka & film & prosecutor  & &  battered   &  kherson   &     mayor&& menacing& connection  &  mire  \\ 
   19 & rapist & appear  & incident   &&  izyum   &    uaf     &  mykolayiv&&  discourage   & breakaway  &  regret  \\ 
   20  & irpin & lie  & protester   &&  mykolaiv   & battalion    &  deport && thereby&  respond    &  gear \\ 
\end{tabular}
\caption{\label{tab:word2vec} Top twenty most similar  terms (by cosine similarity) in each news ecosystem to ``Bucha'', ``Mariupol'',  and ``Ukraine.''  }  
\vspace{-10pt}
\end{table*}
\vspace{2pt}
\noindent 
\textbf{Dissecting Shared Coverage with Word2Vec.} 
To further understand the differences in coverage of similar topics/events between the media ecosystems, we now focus on how each discusses three distinct locales: ``Bucha'', ``Mariupol'', and ``Ukraine''. We choose ``Bucha'' and ``Mariupol'' to cover two of the major cities/events in the Russo-Ukrainian War. In early April 2022, mass graves were discovered in Bucha, leading to outcries about Russian atrocities. The Russian government has denied these claims even as Ukrainian officials have claimed that over 8,000 war crimes were committed in Bucha~\cite{Cheng2022}. We choose Mariupol given its pivotal role in the Russo-Ukrainian War. At the time of writing (May 2022), Mariupol became the epicenter of a humanitarian disaster~\cite{Kirby2022}. We finally choose ``Ukraine''  as ``Ukraine'' covers attitudes of each ecosystem towards the country and the war generally. 

To uncover how each ecosystem has covered our three locales, we build word2vec models. Word2vec models generate word embeddings where words that share similar contexts have parallel vectors in their respective dimensional spaces~\cite{mikolov2013efficient}. We thus utilize them to uncover which words were associated with our three chosen locales. We build a distinct word2vec model for each media ecosystem; we use a training window of seven words and lemmatize and remove stop words~\cite{zannettou2020quantitative}.

\textit{Bucha.} All ecosystems associate  ``Bucha'' with ``atrocity'' and ``killing'' (Table~\ref{tab:word2vec}). The Western ecosystem describes it as ``unspeakable'', ``grisly'', and ``horrifying.'' The Chinese ecosystem further associates the incident as being ``disturbing''. In contrast, the Russian ecosystem associates this town with ``fake'', ``staged'', and ``lie.'' We thus observe that Russian news outlets have echoed the Russian government's denial of war crimes in  Bucha.\footnote{\url{https://web.archive.org/web/20220506041028/https://www.rt.com/russia/553293-bucha-war-crimes-truth/}} However, we also see the use of the words ``alleged'' and ``staged'' in both the Russian and Chinese ecosystems, exhibiting that \textit{both} ecosystems have articles that question whether atrocities even occurred in Bucha~\cite{McCarthy2022}. 

\textit{Mariupol.}
The Western media describes Mariupol as ``besieged'', ``devastated'', and ``battered'' (Table~\ref{tab:word2vec}). This coverage is mirrored by the Chinese media that describes the ``shelling'' of the city. We note the term most associated with Mariupol in both the Russian and Chinese ecosystems (10th in the Western media dataset) is ``Volnovakha'', a city that was leveled in the initial Russian invasion of Ukraine~\cite{Polityuk2022}. This illustrates the degree to which all ecosystems relate Mariupol with the devastation in Volnovakha. We further see that Russian outlets describe the need to ``liberate'' the city and the presence of the ``Azov'' military battalion in the city (Table~\ref{tab:word2vec}). The Azov military battalion was a paramilitary group launched by the Ukrainian ultranationalist groups ``Patriot of Ukraine'' and ``Social-National Assembly'' in 2014. Azov was considered a neo-nazi organization and it was often referenced as a justification for Russia's invasion of Ukraine to ``denazify'' the country~\cite{Thompson2022}. However, we note that despite the Russian call to ``denazify'' Ukraine by ridding it of Azov, this has largely been labeled an attempt to delegitimize Ukrainian interests ~\cite{Thompson2022}. After being reorganized under the National Guard of Ukraine and additional efforts in 2017, the Azov battalion has been largely considered  depoliticized~\cite{Shekhovtsov2020}.

\begin{table*}
\centering
\centering
\fontsize{9pt}{10.5pt}
\selectfont
\setlength{\tabcolsep}{4pt}
\begin{tabular}{clrlrlr}
\toprule \multirow{2}{*}{}  & \multicolumn{2}{c}{\textbf{Western}}  & \multicolumn{2}{c}{\textbf{Russian}}& \multicolumn{2}{c}{\textbf{Chinese}}\\ 
   &{Keywords} & {Articles} & {Keywords}  & {Articles}   & {Keywords}  & {Articles}            \\  \midrule 
   1  & bodies, kyiv, bucha, mariupol,& 141 (3.1\%) & 
     missile,mercenaries,  &  65 (1.8\%)  & talks,round,delegations, &  190   (5.8\%)        \\ 
   
  & civilians,residents,dead &  &
   konashenkov,destroyed,armed &   &
   negotiations,conversation &  \\\cline{2-7}
   
  2  &mariupol,forces,donbas, & 93 (2.1\%) & biological,chemical, &   54  (1.5\%) &china,wang,crisis,& 123 (3.8\%)  \\ 
  
  & official,city,troops & & 
  laboratories,kirillov,nuland & &
  issue,peace,parties & \\ \cline{2-7}
  
   3  & city,building,residential, & 93 (2.1\%)   & azov,nazi,neo-nazi, &  52 (1.5\%)  & biological,laboratories,labs, & 103  (3.2\%)         \\ 
   
  &  killed,buildings,mariupol & & 
  battalion,nazis,neo-nazis && 
  pathogens,zhao,research & \\\cline{2-7}
   
  4& family,daughter,home, &87 (1.9\%) & bucha,kiev,prisoners, &41 (1.4\%)  & parties,china,security, & 73 (2.2\%) \\

   &  husband,friend,mother & & 
  media,videos,lying & & 
  representative,permanent,relevant & \\ \cline{2-7}
  
   5  & children,students,& 86  (1.9\%) &kiev,donbass,zelensky, &  39  (1.1\%)  &crisis,china,security, & 67     (2.1\%)     \\ 
   
   & train,family,parents &  & 
   coup,republics,party & & 
   cold,interests,peace & \\
 
\end{tabular}
\caption{\label{tab:different-topics} The top five (in terms of the number of articles) discriminating topics for each media ecosystem.}  
\vspace{-10pt}
\end{table*}

\textit{Ukraine.} With regards to Ukraine, Western media describes the war as a ``monthslong'' and ``enduring'' conflict (Table~\ref{tab:word2vec}). We further see the Western press write about ``rearming'' Ukraine. For the Russian media, we see the words ``denazify'', and ``demilitarize.'' This echoes the stated aims of the Russian government for invading Ukraine. The Chinese media ecosystem describes the conflict as ``dangerous'' and ``regretful''. This mirrors the statement in early March of Foreign Minister Wang Yi that the Chinese government ``deeply regretted'' the conflict's escalation.\footnote{\url{https://web.archive.org/web/20220514013338/https://www.cnn.com/europe/live-news/ukraine-russia-putin-news-03-01-22/h_ea860211be12bd3bc610264e9922cb17}}

\subsection{Media Ecosystems' Distinctive Topics}
\begin{table}
\centering
\fontsize{9pt}{8.5pt}
\selectfont
\setlength{\tabcolsep}{4pt}
\begin{tabular}{lrlrlr}
\toprule
\multicolumn{2}{c}{\textbf{Western}} & \multicolumn{2}{c}{\textbf{Russian}}& \multicolumn{2}{c}{\textbf{Chinese}} \\ \cmidrule(lr){1-2} \cmidrule(lr){3-4} \cmidrule(lr){5-6}
{Term} & {NPMI} & {Term}  & {NPMI}   & {Term}  & {NPMI}            \\ \midrule
  kyiv  & 0.1001 &   kiev &  0.1291	&  china     & 0.1222	\\  
  invasion  & 0.0990&    donbass &  0.1183&   wang   &0.1193	 \\  
   russians  &0.0930 &   dpr &  0.1126	  &  councilor  & 0.1064	 \\ 
zelenskyy  & 0.0876&   	lpr & 0.1072  &  sustainable    & 0.1034  	\\ 
  zelenskiy  & 0.0853&    republic &0.1062 	     &  zhang  & 0.0947  		 \\ 
 separatist  & 0.0835  &  afu &   0.1011	    & percent   &0.0945	\\ 
 listen  &0.0824&     lugansk & 0.0991	    &  zhao  & 0.0880	  \\ 
   official  & 0.0806 &   neo-nazi &0.0892	& growth   & 0.0853\\ 
   mile  & 0.0791 &   	  coup &  0.0871   &   conducive  & 0.0828	 \\ 
  ukrainians  & 0.0778 &    nationalist &  0.0856    &  dialogue &  0.0817	\\ 
\end{tabular}
\caption{\label{tab:characteritic-words}The most characteristic words for each media ecosystem. }  
\vspace{-10pt}
\end{table}
Having explored the shared topics discussed by each ecosystem and how each ecosystem discussed these topics in a different manner, we now using PLDA and our NPMI metric seek to uncover the \textit{distinctive} topics and words that each media ecosystem emphasizes in articles published between January 1 and April 15, 2022.

\vspace{2pt}
\noindent 
\textbf{Western Media Coverage.} Using PLDA, we see that the Western press focuses thoroughly on the military aspects of the Russo-Ukrainian conflict (Table~\ref{tab:different-topics}). The top three most discriminating topics discovered through PLDA all discuss military aspects of the war and the destruction within Ukraine with words like ``dead'', ``killed'', and ``bodies.'' We further see that the Western press has focused significantly on crimes committed in Ukraine and the plight of Ukrainian refugees. The most discriminating topic specifically concerns the city of Bucha where mass graves were discovered in early April~\cite{Cheng2022}.

In contrast to the larger themes particular to the Western media, using our NPMI metric, we see the most characteristic words of the Western ecosystem have to do the with seemingly innocuous spelling of Ukrainian cities (\textit{e.g.}, Kyiv) and the president of Ukraine, Volodymyr Zelenskyy's, name (Table~\ref{tab:characteritic-words}). We note, however, that the spelling of these entities has taken on political meaning~\cite{Dickinson2019}. Different ecosystems, in their reporting on events in Ukraine, have decided either to utilize the Ukrainian or Russian spelling of words. Here we see that the Western ecosystem, in contrast to both the Russian and Chinese ones, utilizes Ukrainian spellings. We further see this mirrored in the most uncharacteristic words within this dataset (Table~\ref{tab:uncharacteristic-words}), where we see ``Kiev, Lugansk, and Kharkov'' rather than their Ukrainian spellings of ``Kyiv, Luhansk, and Kharkiv.''

\begin{table}
\centering
\fontsize{9pt}{8.5pt}
\selectfont
\setlength{\tabcolsep}{4pt}
\begin{tabular}{lrlrlr}

\toprule
\multicolumn{2}{c}{\textbf{Western}} & \multicolumn{2}{c}{\textbf{Russian}}& \multicolumn{2}{c}{\textbf{Chinese}} \\ \cmidrule(lr){1-2} \cmidrule(lr){3-4} \cmidrule(lr){5-6}
{Term} & {NPMI} & {Term}  & {NPMI}   & {Term}  & {NPMI}            \\ \midrule
  kiev  & -0.2508&   kyiv & -0.2016	&   invasion   &-0.1821		\\ 
  donbass  &-0.1546&    zelenskyy & -0.1910 		  &    zelenskiy  &-0.1794 \\
  dpr  & -0.1528 &   zelenskiy &  -0.1348		  &  russians     & -0.1628		 \\  
  lugansk  & -0.1522&   percent & -0.1324		  &    defence  & -0.1506		 \\ 
  wang  & -0.1412&    wang & 	-0.1311	      &   nazi   &	-0.1466		 \\ 
  lpr  & -0.1405&  shelter &   -0.1220		    & ukrainians  & -0.1443	\\
  afu  & -0.1389 &     flee & -0.1210		    &  crime    &	-0.1373	  \\ 
  zhang  & -0.1211&    lviv & -0.1176		 & soldier   & -0.1361 	 \\ 
  kharkov  &-0.1102 &   listen &  -0.1174	     &   attack  &  -0.1357		 \\ 
  councilor  & -0.1017 &     invasion & 	-0.1167	     &  strike &  -0.1351		\\ 
\end{tabular}
\caption{\label{tab:uncharacteristic-words} The most uncharacteristic words for each media ecosystem.}  
\vspace{-10pt}
\end{table}

\vspace{2pt}
\noindent 
\textbf{Russian Media Coverage.}
In contrast to Western media, Russian media articles focus on the Donbas region of Ukraine and the portrayal of the Ukrainian government and its military (AFU) as being filled with ``neo-nazis.'' Notably, as seen in Table~\ref{tab:different-topics}, Russian media (39~articles) have a keen focus on the Donbas region of Ukraine, which is partly controlled by two separatist regimes: the Donetsk People's Republic (DPR) and the Luhansk People's Republic (LPR). While still parts of Ukraine, significant efforts from the Russian media ecosystem have gone into demonstrating a distinction between these area's regimes and internationally recognized Ukrainian government.\footnote{\url{https://web.archive.org/web/20220501202030/https://www.rt.com/russia/552285-donbass-russia-ukraine-history/}} The day before the Russian invasion of Ukraine, the leaders of DPR and LPR called for Russia to help ``repel Kiev's aggression'', further ostensibly prompting the Russian government to invade and ``liberate'' the regions.\footnote{\url{https://web.archive.org/web/20220223210219/https://tass.com/politics/1409091}} With Russian media recognizing these regimes, the DPR and LPR are some of the most characteristic words in the Russian news ecosystem; and simultaneously the least descriptive of the Western news ecosystem  (Tables~\ref{tab:characteritic-words} and~\ref{tab:uncharacteristic-words}).

We further observe a focus on Ukraine being filled with ``neo-nazis''. The word ``neo-nazi'' is one of the most characteristic words utilized by Russian media (Table~\ref{tab:characteritic-words}) and a part of the most discriminating Russian topic(Table~\ref{tab:different-topics}). As previously noted, a major aspect of the Russian claim that Ukraine required ``denazification'' was that the Azov battalion was a part of Ukraine's military; (again) this despite Azov's depoliticization and Ukraine's low level of antisemitism~\cite{Shekhovtsov2020,Masci2018}. Ukraine's current president Volodymyr Zelenskyy is even Jewish~\cite{Thompson2022}.

Lastly, we note the disinformation narrative about biological weapons that appeared as the third discriminating topic for Russian articles. On March 6, 2022, the Russian news reporting website Tass reported that the US had been funding biological weapons development within Ukraine and had subsequently destroyed the facilities upon the Russian invasion. This narrative was debunked by the US State Department~\cite {Price2022} and the New York Times~\cite{Qiu2022}.

\vspace{2pt}
\noindent 
\textbf{Chinese Media Coverage.}
As seen in Table~\ref{tab:different-topics} and Table~\ref{tab:characteritic-words}, the most discriminating aspects of the Chinese media are its concerns with the economic and diplomatic fallout of the war in Ukraine. The largest discriminating topic that was uncovered by PLDA concerned ``negotiations'' and attempts to broker peace. We see this echoed further in the second, fourth, and fifth most discriminating topics. This largely matches the view found through the sentiment analysis approach in Section~\ref{sec:compare_text}.  

In addition to emphasizing ``negotiations'', we further observe a heavy emphasis on the economic ramifications of the Ukraine crisis on the economic health of China and the rest of the globe. Several of the most distinctive words, and the sixth most discriminating topic discovered (not shown), discuss primarily the effect that the Russo-Ukrainian conflict will have on food prices. Because Russia and Ukraine together comprise about 30\% of global wheat exports and 20\% of maize exports, the war has caused major disruptions to commodity prices. The United National Food and Agricultural Organization found that food commodity prices went up 12.6\% between February and March of 2022~\cite{Treisman2022}. This particular worry about food prices was echoed widely in the Chinese news ecosystem with a word like ``percent'', ``growth'', and ``sustainable'' being characteristic of its news reporting on Ukraine.

We additionally note that the Russian disinformation narrative of US-funded Ukrainian biological laboratories found significant traction within Chinese media. The Chinese spokesman for the Foreign Ministry even tweeted a video directly from Russia Today about these supposed biological weapons laboratories\footnote{\url{https://web.archive.org/web/20220501204212/https://twitter.com/zlj517/status/1501758675006148610}} and we see this largely throughout the Chinese news ecosystem (Table~\ref{tab:different-topics}) with 103 articles.

\begin{CJK*}{UTF8}{gkai}
\begin{table*}
\centering
\fontsize{9.0pt}{8.5pt}
\selectfont
\setlength{\tabcolsep}{4pt}
\begin{tabular}{lrr|lrr|lr}
\toprule
\multicolumn{3}{c}{\textbf{Weibo}} & \multicolumn{3}{c}{\textbf{Twitter}} & \multicolumn{2}{c}{\textbf{News Articles}}\\ \midrule

  & { Russian } & {} & & {Russian }  & {} & & {Russian }\\ 
 User& Citations &Russian ``\#'' &User  & Citations & Russian ``\#'' & Domain&  Citations
 \\\midrule
 
中国新闻网 (China News Service) & 142 & 63 &  Echinanews & 36 & 804  &ecns.cn & 26 \\
中国日报 (China Daily) & 65 & 76 & ChinaDaily & 12 & 2178 &chinadaily.com.cn & \textbf{68} \\
CGTN/CGTN记者团 & 69 & \textbf{87} & CGTNOfficial& \textbf{468} & \textbf{3487}  &cgtn.com & 45   \\
人民日报 (People's Daily)&  6 & 20 &  PDChina & 65  & 826  &pdnews.cn & 14\\ 
央视新闻 (CCTV)  & 11  & 68 &  CCTV & 3 & 34  &cctv.com & 6\\ 
环球时报 (Global Times) &\textbf{395}& 83 & globaltimesnews  &33  & 2729&globaltimes.cn & 27\\
新华网 (Xinhua News) &35& 27 & XHNews  &21  &804& xinhuanet.com & 35\\ 
\end{tabular}
\caption{\label{tab:chinese-gov-reposting-per-user} The number of posts for each Chinese outlet per platform referencing a Russian news organization as a news source or posting a hashtag utilized by Russian state-sponsored accounts between January 1 and April 15, 2022.}  
\vspace{-10pt}
\end{table*}
\end{CJK*}

\subsection{Summary}
The Western, Russian, and Chinese news ecosystems have largely utilized divergent narratives in covering the Russo-Ukrainian war. In describing the invasion, Western media has utilized Ukrainian spellings of Ukrainian towns and officials and have highlighted the military and humanitarian crises taking place in Ukraine. Chinese outlets have avoided describing some of the more grisly aspects of the invasion, referring to Russo-Ukrainian War as a ``conflict'' and ``crisis'', causing the overall sentiment score of their articles to be somewhat higher than the scores of Russian the Western media. Finally, the Russian media ecosystem has largely focused on the purported reasons for the ``special military operation'', ``the denazification'' of Ukraine, and the supposed ``liberation'' of Ukraine's people.

\section{Russian Influence on Chinese State Media}
Despite differences in the topics discussed by Chinese and Russian ecosystems, we observed that Chinese media outlets echoed narratives from Russian news sources. In particular, Chinese news outlets amplified the disinformation narrative that the US had funded Ukraine-based biological weapons laboratories. We saw a similar Russian influence in the coverage of the massacre at Bucha (Table~\ref{tab:word2vec}). In this section, we document the degree to which Russian state media have influenced Chinese media coverage and discussions of the Russo-Ukrainian War in their news articles, on Weibo, and on Twitter. We take a multi-modal approach, quantifying influence along two axes: (1) citations of Russian outlets as news sources and (2) reuse of Russian-sourced images.

\begin{figure}
\begin{subfigure}{.48\textwidth}
  \centering
  \includegraphics[width=1\linewidth]{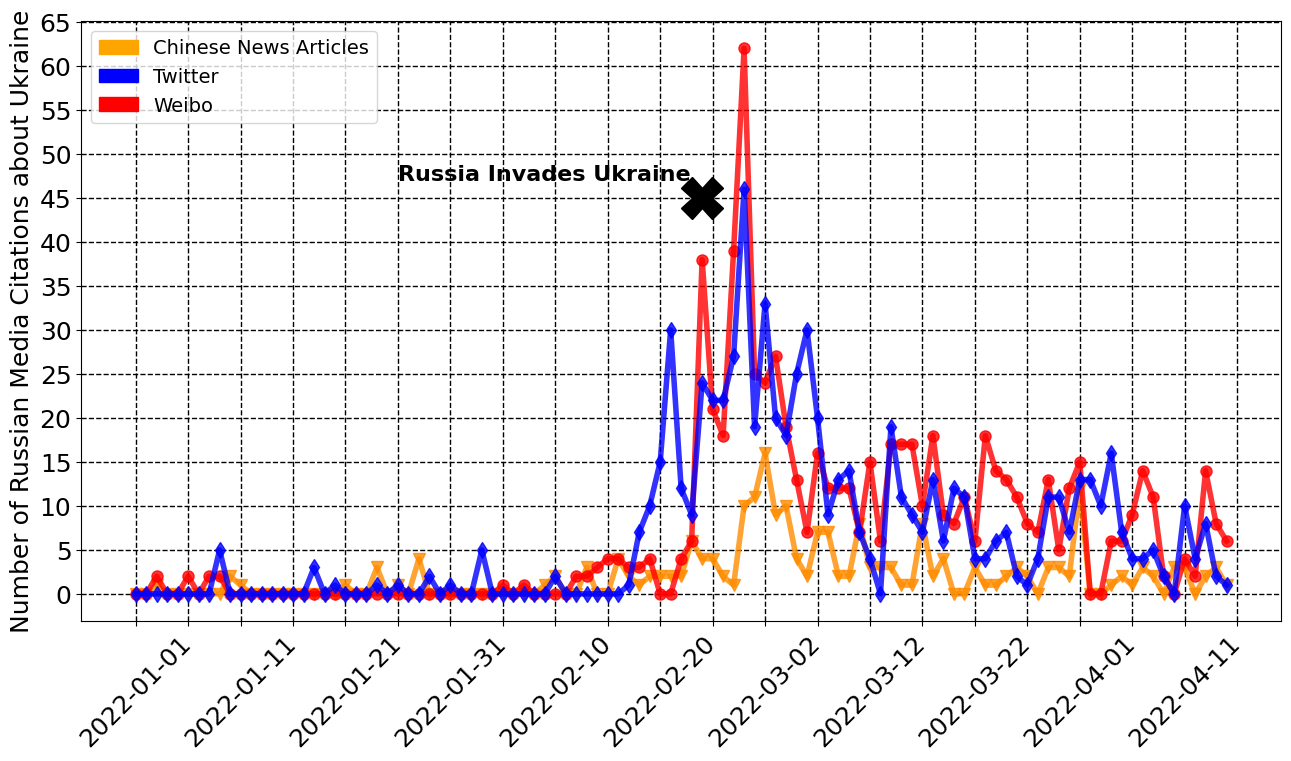}
\label{fig:rt}
\end{subfigure}%
\caption{Citations of Russian news sources (Tass, RIA, Sputnik News, and RT) by Chinese media outlets. 
}
\label{fig:china-russia-shared-gov-picture}
\vspace{-10pt}
\end{figure}

\subsection{Chinese Media's Use of Russian News Sources}
We first examine citations of Russian news sources within Chinese state media reporting. Specifically, we look at Chinese media's citations of the Russian state-controlled news agencies Tass, Russia Today, Sputnik News, and RIA Novosti (the prominent news agency that started Sputnik News~\cite{Feinberg2017}) as sources for reporting about Ukraine across online news articles, on Weibo, and on Twitter. As an example of what we consider a citation, on May 13th, the Twitter account of CGTN tweeted: \textit{Leader of South Ossetia sets referendum on joining Russia on July 17, TASS reported}. For each mention of the Russian outlets across each platform, we manually verify that the Chinese state outlet utilized them as a news source rather than just mentioning the outlet (for our Weibo, one of the authors can read Mandarin Chinese).


The number of citations of Russian news sources in Chinese news reporting on Ukraine saw a sharp uptick beginning in February (Figure~\ref{fig:china-russia-shared-gov-picture}). We note that this peak occurred across all platforms. We further note that across every platform, each Chinese news source utilized a Russian news organization at least once in their reporting about Ukraine (Table~\ref{tab:chinese-gov-reposting-per-user}). Global Times on Weibo, CGTN on Twitter, and China Daily within their news articles most consistently relied on Russian news sources. As seen in Table~\ref{tab:chinese-gov-reposting-perdomain}, this is particularly true for news sources like Tass and RIA Novosti, which are consistently utilized across different platforms. Looking at Twitter and Weibo, every outlet similarly also extensively tweeted and posted to Weibo utilizing hashtags utilized by Russian state-sponsored accounts (Sputnik News, Russia Today, and the Russian Embassy).

\begin{CJK*}{UTF8}{gkai}

Looking at our extended set of users on Weibo, beyond our core set of Chinese media outlets and Russian entities (RT, Russian Embassy, and Sputnik News), we identify another 4,478 references to Russian news outlets among the remaining users. Among these users are several other Chinese news outlets: Reference News/参考消息 585 times, Sina News/新浪军事 369 times, and Observer News/观察者网 310 times. These same outlets utilize hashtags on Weibo shared by RT, Russian Embassy, and Sputnik News 96, 113, and 75 times within the same period respectively (the 2nd, 1st, and 5th most often in our dataset respectively). Overall, across every non-Russian Weibo account (188 accounts), we see that 132 (70.2\%) reference a Russian news outlet, and all use hashtags also shared by Russian state-sponsored Weibo accounts. Collectively, the posts that cited Russian sources received 10,792,782 reactions (Weibo version of likes), with a median of 114 reactions per post. As an aside, the most popular post (579,146 reactions), submitted by Sina News/新浪军事\ and citing the RIA news agency, described the destruction of Ukrainian ships in Odesa.
\end{CJK*}


\subsection{Chinese Media’s Reuse of Russian-sourced Images}
To understand the degree to which Russian-sourced imagery has influenced the Chinese media and users, we determine what percentage of their \emph{Ukraine-related} images came from Russian sources. We utilize the Google Cloud Vision API's reverse search functionality to determine the set of an image's possible sources.\footnote{https://cloud.google.com/vision/} Returned with this search are the entities associated with the photo and a list of other website pages that posted an identical image. We consider an image to be \emph{Ukraine-related} if one of the entities returned with the image is ``Ukraine.'' We consider a Ukrainian-related image to be of Russian origin if it was first posted on a .ru website or one of our Russian news websites \textit{prior} to when it was posted on Weibo. We visit each page returned with our reverse-image search with  Selenium and utilize the Python library \texttt{htmldate} to acquire the publish date to ensure that the image was posted on a Russian page first.
\begin{table}
\centering
\centering
\fontsize{9.0pt}{8.5pt}
\selectfont
\setlength{\tabcolsep}{4pt}
\begin{tabular}{lrrr}
\toprule
Russian News Outlet &\textbf{Weibo} & \textbf{Twitter} & \textbf{News Articles}\\\midrule
RIA Novosti &189 &\textbf{261}  &60 \\
Tass &203  & 206 & \textbf{96}\\
RT &\textbf{223} &56  & 24\\ 
Sputnik News& 175  &132  & 40 \\ 
\end{tabular}
\caption{\label{tab:chinese-gov-reposting-perdomain} Citations by Chinese news outlets to  Russian news sources between January 1 and April 15, 2022.}  
\vspace{-10pt}
\end{table}


\begin{CJK*}{UTF8}{gkai}
We first perform a reverse image search on the images posted within the tweets and Weibo posts of our set of seven Chinese state media outlets. Altogether this consists of 9,393~images across 29,052~Weibo posts and 7,809~images across 44,804~unique tweets. After running our set of Weibo images through Google's Cloud API, we identify only 18 (0.2\%) different images that were of Russian origin, 6 of them coming from rt.com and 4 from sputniknews.com. 10 of these images were posted by Global Times/环球时报, 4 by CGTN, 2 from People's Daily, 1 from China Daily, and 1 from CCTV. Among the 7,809 images posted by Chinese state media Twitter accounts, only 6 (0.07\%) photos were of Russian origin. For Chinese state media, we thus see that they largely did not utilize imagery from Russian sources. 
\end{CJK*}


While Chinese media outlets did not heavily utilize Russian-sourced/associated images, we \emph{do} see elevated usage of Russian-sourced imagery amongst the most prominent Ukraine-related Weibo accounts. Of the 96,614~images posed by these accounts between January~1, 2022, and April~15, 2022, 13,854 (14.3\%) of images were \textit{Ukraine-related}, and 1,196 (8.6\%) of images came directly from a Russian source. Similarly, among the 16,063 images about Russia, 2,313 (14.4\%) of these images came from Russian sources.  For images not about Ukraine nor Russia, only 1.2\% of images were from Russian sources. 109 (58\%) of the 188 accounts not directly associated with Russia, posted at least one Russian-sourced image about Ukraine, with~14~(7.4\%) users posting a higher percentage of Russian-sourced images than the Russian embassy account (21\%).


\subsection{Russian Disinformation Narratives on Weibo} 
\begin{CJK*}{UTF8}{gkai}
\begin{table}
\centering
\centering
\fontsize{9.0pt}{8.5pt}
\selectfont
\setlength{\tabcolsep}{4pt}
\begin{tabular}{lrr}
\toprule
 &   \textbf{Ukrainian} 
& \textbf{Ukrainian} \\ \
Account& \textbf{Bio-Weapons Posts}  & \textbf{Nazis Posts} \\ \midrule
中国新闻网 & 20 & 3 \\
中国日报  & 138 &10  \\\
CGTN/记者团 & 44 &  24  \\
人民日报& 109 & 3 \\
央视新闻  & 121 & 0   \\ 
环球时报 &\textbf{165} & \textbf{96} \\ 
新华网 &25& 1  \\
\end{tabular}
\caption{\label{tab:chinese-gov-reposting-disinformation} References to Russian disinformation campaigns. 
}
\vspace{-10pt}
\end{table}

\end{CJK*}
\begin{CJK*}{UTF8}{gkai}
Having shown the increased influence of Russian news sources within online Chinese communities, we finally look at if this increased influence led to the increased spread of Russian disinformation narratives by Chinese news outlets.

As shown in the previous section, outlets like the Global Times have a higher level of connection to Russian outlets than others like CCTV. While conducting our previous analyses, we observed that these same outlets often posted about the Russian disinformation campaign about US-funded Ukrainian biological weapons laboratories and Ukrainian neo-nazis (Table~\ref{tab:chinese-gov-reposting-disinformation}). For example, 环球时报/Global Times had 261~different Weibo posts concerning these stories (165~about Ukrainian biological weapons and 96 about Ukrainian neo-nazis) while also referencing Russian sources 415~times and posting 10~Russian-sourced images. Similarly, the news outlet with the most posts about these disinformation narratives (182 about biological weapons, and 87~about Ukrainian neo-nazis) Sina News/新浪军事 also had elevated amounts of references to Russian outlets as sources. In this section, we thus model the connection of different Chinese news outlets to the Russian government and Russian state media and their tendency to publish Russian disinformation stories. 

\vspace{2pt}
\noindent 
\textbf{Experimental Setup.} For this experiment, we utilize our extended Weibo dataset, including the additional 32 Chinese state media entities that were in Fung et al.'s dataset; altogether 39~outlets.  To model each news outlet's promotion of Russian disinformation about Ukraine, we use the combined count of their mentions of the US-funded Ukrainian biological weapons laboratories and their mentions of the Ukrainian government being dominated by neo-nazis. We utilize these two disinformation narratives as they were heavily promoted by Russian websites as seen in Section~\ref{sec:compare_text}. Furthermore, among our 39~Chinese outlets, posts that mentioned these disinformation narratives were fairly popular (the posts for each collectively received 2,756,638 and 409,368 Weibo reactions respectively). Running Mann-Witney U-tests comparing the popularity of posts (based on the number of reactions) mentioning these stories to posts not mentioning either, we find in both cases that these stories enjoyed higher popularity than posts not mentioning them. Posts mentioning these two narratives received an average of 3,132 reactions and 1,347 reactions respectively.\end{CJK*} 

To properly model the interplay between our variables indicating an outlet's connection to Russian entities and the counts of its publication of Russian disinformation, we utilize a negative binomial regression. We utilize a negative binomial rather than Poisson regression for data given that negative binomial models do not make the strict assumption that the mean of the data is equivalent to the variance; this assumption has been shown to be unrealistic for many real-world scenarios~\cite{morina2022web}. As input variables, we utilize the number of citations of Russian news agencies, the number of shared images from Russian sources, and finally the number of shared hashtags with the Weibo version of RT, Sputnik News, and the Russian Embassy. Looking initially at these variables we see that have 0.716, 0.600, and 0.671 Pearson correlations respectively with the number of Russian disinformation stories that each individual Chinese outlet published. This already illustrates that outlets with close ties to Russian entities tend to post more Russian disinformation narratives. Finally, we note that we regularize disinformation story counts by the number of posts each account made during our collection period.

\begin{table}
\centering
\fontsize{9.0pt}{8.5pt}
\selectfont
\setlength{\tabcolsep}{4pt}
\begin{tabular}{lrr}
\toprule
\textbf{} & \textbf{Coefficient} & \textbf{p-value} \\ \midrule
Intercept   & 1.495 & 0.000\\ 
\# Russian-sourced & 0.0163 & 0.013 \\
\# Russian photos  & -0.0275	&0.439	  \\ 
\# Russian hashtags & 0.0533 &  0.393\\ 
\end{tabular}
\caption{\label{tab:neg-binomial} Fit of the negative binomial regression.
}
\vspace{-10pt}
\end{table}

\vspace{2pt}
\noindent 
\textbf{Results.} As seen in Table~\ref{tab:neg-binomial}, Russian news source utilization \textit{is} a predictor of posting about Russian disinformation with a coefficient of 0.0168 (1 more Russian citation leads to an expected 1.6\% increase in Russian disinformation stories [\textit{e.g.}, for every 42 Russian news citation, the number of expected disinformation stories doubles]). With a p-value cutoff of 0.05, we see that it is the only one of our variables that is a predictor of Chinese state media accounts' usage of Russian disinformation stories. The other two variables that we considered were largely accounted for after taking into account how often a given account cited Russian news outlets. We thus see that the increased influence of Russian news on Chinese outlets \textit{is} correlated with increased amounts of Russian disinformation stories within Chines media ecosystems.



\section{Related Work}
We are not the first to perform a differential analysis of different text corpora to understand their treatment of given topics. Olteanu et al.~\cite{olteanu2015comparing} compare differential discourse surrounding climate change. Similarly, Galvez et al.~\cite{galvez2019half} compare the differential word associations of different genders.


Our work further builds several others' in quantifying the use of images in the spread of narratives. One of the first to take an image-based approach in examining the contents of social media misinformation, Seo et al. characterized the use of images as propaganda during the 2012 Israeli--Hamas conflict~\cite{seo2014visual}. More recently, Zannettou et al. found that Russian state-sponsored Twitter accounts had a marked influence on the memes or images that appeared on Reddit and the alternative social media platform Gab~\cite{zannettou2018origins}. Like our approach that considers both images, metadata, and text of Russian origin,  Zhang et al.~\cite{zhang2019multi} utilize a multi-modal approach that takes into account both images and text to track misinformation and rumors on Twitter. By looking at images and hyperlinks, in their papers, Wilson and Starbird et al. look at the influence of Syrian White Helmets across different platforms and Hanley et al. look at the spread of QAnon~\cite{wilson2020cross,hanley2021no}. 


Similar to our work, studies in the past decade, in particular, have studied several Russian disinformation and propaganda campaigns aimed at sowing division~\cite{badawy2018analyzing}. For example, Diresta et al. found that during the 2016 US election, 1852 different Facebook and Instagram ad campaigns were targeted at ethnic minority communities~\cite{diresta2019tactics}. In the years following the 2016 election, Badawy et al. found that Russian bots parroted US pro-conservative and divisive messages. They further found that most of these messages were specifically targeted at people in the Southern US states~\cite{badawy2018analyzing}. Outside of the US, many of these Russian bots and troll accounts have also spread Russian propaganda and disinformation about Ukraine since Russia's illegal annexation of Crimea within Eastern Europe~\cite{hellman2017can}.  However, it is not only Russian state-sponsored accounts that have spread Russian propaganda online. Golovchenko et al. found a larger majority of the message on Twitter promoting pro-Russian narratives surrounding events in Ukraine belonged to personal Russian non-state-sponsored accounts~\cite{golovchenko2018state}. As a whole, 1,811 different Twitter profiles belonging to individual users rather than to large state media or journalist accounts had the biggest effect in generating pro-Russian propaganda in their study. 




\section{Discussion and Conclusion }
In this work, we employ a quantitative approach to understand the distinctive narratives about the Russo-Ukrainian War being spread by Western, Russian, and Chinese media. Utilizing PLDA, the NMPI information metric, word2vec models, and differential sentiment analysis, we find that each ecosystem, in turn, has entirely different and nuanced perspectives. Looking at the topics written and performing a cross-platform study, we show a spike in the influence of Russian news outlets on the Chinese news ecosystem since the beginning of the Russo-Ukrainian War.

When conducting such studies, we argue that uncovering \textit{why} the overall sentiment and coverage of different ecosystems is different is pivotal.  For example, we found that the reason why Chinese ecosystems had a more positive sentiment about the war is due to their emphasis on cooperation and peace negotiations. Similarly, our finding that Russian and Chinese outlets often refer to the war as a ``conflict'' or ``crisis'' while the Western press refers to it as a ``war'' and ``invasion'' colors \textit{why} different ecosystems have distinctive perspectives on the ongoing military operations in Ukraine. Only by taking an approach that incorporates several methods (\textit{e.g.}, PLDA, word2vec) can these insights be uncovered.

Lastly, we note that it is imperative to acknowledge that events like the Russo-Ukrainian War are global with no country in the world not affected~\cite{Treisman2022}. Given the highly polarized and global nature of the Russo-Ukrainian War, understanding how different media ecosystems are covering it is imperative to collectively understand broader perceptions of the conflict. Previous studies that have focused only on news targeted at a Western audience 
are likely limited and may experience bias. We urge researchers to work on tracking and understanding how these narratives spread to account for non-Western and non-English language audiences. The Russian disinformation website waronfakes.com for example has versions in English, French, Dutch, Spanish, Arabic, and Chinese.Without taking into account how the influence of Russian news on the Chinese media ecosystem, for example, would have largely ignored a key piece to understanding instances of pro-Russian bias within Chinese news coverage.

\vspace{2pt}
\noindent
\textbf{Future Work.} Using our approach we described how Chinese, Russian, and Western press reported on military operations in Ukraine. Our approach can be naturally extended to understand differential coverage of events beyond the Russo-Ukrainian War and to other countries' news ecosystems. While we focus on the war here due to its importance, we propose utilizing differential sentiment analysis, PLDA, and NMPI to uncover differing perspectives across many different news stories.  

\noindent
\textbf{Ethical Considerations.}
Within this work, we utilize public data and follow ethical guidelines as outlined by others ~\cite{hanley2021no}. We do not seek to deanonymize users within our Weibo and Twitter datasets. We recognize that Russo-Ukrainian War is an ongoing conflict and that information about the war is changing day to day. We hope to remain objective and sensitive about the issues discussed here.
\section*{Acknowledgements} 
This work was supported in part by the National Science Foundation under grant \#2030859 to the Computing Research Association for the CIFellows Project, a gift from Google, Inc., and NSF Graduate Fellowship DGE-1656518. 
\bibliography{paper}

\end{document}